\documentclass[onecolumn]{mn2e}  
\usepackage[dvips]{graphics}
\usepackage{epsfig}
\usepackage{epsf}
\usepackage[figuresleft]{rotating}
\begin{document}
%
\newcommand{\getinptfile}
{\typein[\inptfile]{Input file name}
\input{\inptfile}}

\newcommand{\mysummary}[2]{\noi {\bf SUMMARY}#1 \\ \noi \sl #2 \\ \capline 
	\hspace{-.13in} \raisebox{.0in}{$\sqcap$} \rm }  
\newcommand{\mycaption}[2]{\caption[#1]{\footnotesize #2}} 
\newcommand{\capline}{\mbox{}\hrulefill}
\newcommand{\mysection}[2]{ 
\section{\uppercase{\normalsize{\bf #1}}} \def\junksec{{#2}} } %
\newcommand{\mychapter}[2]{ \chapter{#1} \def\junkchap{{#2}}  
\def\thesection{\arabic{chapter}.\arabic{section}}
\def\thesubsection{\thesection.\arabic{subsection}}
\def\thesubsubsection{\thesubsection.\arabic{subsubsection}}
\def\theequation{\arabic{chapter}.\arabic{equation}}
\def\thefigure{\arabic{chapter}.\arabic{figure}}
\def\thetable{\arabic{chapter}.\arabic{table}}
}
\newcommand{\mysubsection}[2]{ \subsection{#1} \def\junksubsec{{#2}} }
\def\thenote{\addtocounter{footnote}{1}$^{\scriptstyle{\arabic{footnote}}}$ }

\newcommand{\myfm}[1]{\mbox{$#1$}}
\def\spose#1{\hbox to 0pt{#1\hss}}	
\def\ltabout{\mathrel{\spose{\lower 3pt\hbox{$\mathchar"218$}} 
     \raise 2.0pt\hbox{$\mathchar"13C$}}}
\def\gtabout{\mathrel{\spose{\lower 3pt\hbox{$\mathchar"218$}}
     \raise 2.0pt\hbox{$\mathchar"13E$}}}
\newcommand{\ltsim}{\raisebox{-0.5ex}{$\;\stackrel{<}{\scriptstyle \backslash}\;$}}
\newcommand{\simlt}{\ltsim}
\newcommand{\simgt}{\gtsim}
%
\newcommand{\unit}[1]{\ifmmode \:\mbox{\rm #1}\else \mbox{#1}\fi}
\newcommand{\ze}{\ifmmode \mbox{z=0}\else \mbox{$z=0$ }\fi }

%
\newcommand{\boldv}[1]{\ifmmode \mbox{\boldmath $ #1$} \else 
 \mbox{\boldmath $#1$} \fi}
%
\renewcommand{\sb}[1]{_{\rm #1}}%
\newcommand{\expec}[1]{\myfm{\left\langle #1 \right\rangle}}
\newcommand{\mone}{\myfm{^{-1}}}
\newcommand{\half}{\myfm{\frac{1}{2}}}
\newcommand{\nth}[1]{\myfm{#1^{\small th}}}
\newcommand{\ten}[1]{\myfm{\times 10^{#1}}}
\newcommand{\abs}[1]{\mid\!\! #1 \!\!\mid}
\newcommand{\as}{a_{\ast}}
\newcommand{\asr}{(a_{\ast}^{2}-R_{\ast}^{2})}
\newcommand{\bvm}{\bv{m}}
\newcommand{\calf}{{\cal F}}
\newcommand{\calI}{{\cal I}}
\newcommand{\calm}{{v/c}}
\newcommand{\calminf}{{(v/c)_{\infty}}}
\newcommand{\calQ}{{\cal Q}}
\newcommand{\calR}{{\cal R}}
\newcommand{\calw}{{\it W}}
\newcommand{\co}{c_{o}}
\newcommand{\cs}{C_{\sigma}}
\newcommand{\cst}{\tilde{C}_{\sigma}}
\newcommand{\cv}{C_{v}}
\def\dbar{{\mathchar '26\mkern-9mud}}	
\newcommand{\deldelr}{\frac{\partial}{\partial r}}
\newcommand{\deldelR}{\frac{\partial}{\partial R}}
\newcommand{\deldeltheta}{\frac{\partial}{\partial \theta} }
\newcommand{\deldelphi}{\frac{\partial}{\partial \phi} }
\newcommand{\ddotrc}{\ddot{R}_{c}}
\newcommand{\ddotxc}{\ddot{x}_{c}}
\newcommand{\dotrc}{\dot{R}_{c}}
\newcommand{\dotxc}{\dot{x}_{c}}
\newcommand{\Estar}{E_{\ast}}
\newcommand{\grpsi}{\Psi_{\ast}^{\prime}}
\newcommand{\kboltz}{k_{\beta}}
\newcommand{\levi}[1]{\epsilon_{#1}}
\newcommand{\limaso}[1]{$#1 ( a_{\ast}\rightarrow 0)\ $}
\newcommand{\limasinfty}[1]{$#1 ( a_{\ast}\rightarrow \infty)\ $}
\newcommand{\limrinfty}[1]{$#1 ( R\rightarrow \infty,t)\ $}
\newcommand{\limro}[1]{$#1 ( R\rightarrow 0,t)\ $}
\newcommand{\limrso}[1]{$#1 (R_{\ast}\rightarrow 0)\ $}
\newcommand{\limxo}[1]{$#1 ( x\rightarrow 0,t)\ $}
\newcommand{\limxso}[1]{$#1 (\xs\rightarrow 0)\ $}
\newcommand{\ls}{l_{\ast}}
\newcommand{\Ls}{L_{\ast}}
\newcommand{\mean}[1]{<#1>}
\newcommand{\ms}{m_{\ast}}
\newcommand{\Ms}{M_{\ast}}
\def\nb{{\sl N}-body }
\def\nbt{{\sf NBODY2} }
\def\nb1{{\sf NBODY1} }
\newcommand{\nuoned}{\nu\sb{1d}}
\newcommand{\ra}{\rightarrow}
\newcommand{\Ra}{\Rightarrow}
\newcommand{\rc}{r_{c} } 
\newcommand{\Rc}{R_{c} } 
\newcommand{\res}[1]{{\rm O}(#1)}
\newcommand{\rnsa}{(r^{2}-a^{2})}
\newcommand{\Rnsa}{(R^{2}-a^{2})}
\newcommand{\rs}{r_{\ast}}
\newcommand{\Rs}{R_{\ast}}
\newcommand{\Rsa}{(R_{\ast}^{2}-a_{\ast}^{2})}
\newcommand{\sa}{\sigma } 
\newcommand{\sac}{\sigma_{c} } 
\newcommand{\sas}{\sigma_{\ast} } 
\newcommand{\sasp}{\sigma^{\prime}_{\ast}}
\newcommand{\saxs}{\sigma_{\ast} } 
\newcommand{\sech}{{\rm sech}}
\newcommand{\tff}{t\sb{ff}} 
\newcommand{\ti}{\tilde}
\newcommand{\trel}{t\sb{rel}}
\newcommand{\ts}{\tilde{\sigma} } 
\newcommand{\tss}{\tilde{\sigma}_{\ast} } 
\newcommand{\vcol}{v\sb{col}}
\newcommand{\vs}{v_{\ast}  } 
\newcommand{\vsp}{v^{\prime}_{\ast}}
\newcommand{\vxs}{v_{\ast}  } 
\newcommand{\xs}{x_{\ast}}
\newcommand{\xc}{x_{c} } 
\newcommand{\xistar}{\xi_{\ast}}
\newcommand{\rmd}{\ifmmode \:\mbox{{\rm d}}\else \mbox{ d}\fi }
\newcommand{\rmD}{\ifmmode \:\mbox{{\rm D}}\else \mbox{ D}\fi }
\newcommand{\valfven}{v_{{\rm Alfv\acute{e}n}}}

%
\newcommand{\noi}{\noindent}
\newcommand{\bc}{boundary condition }
\newcommand{\bcs}{boundary conditions }
\newcommand{\Bcs}{Boundary conditions }
\newcommand{\lhs}{left-hand side }
\newcommand{\rhs}{right-hand side }
\newcommand{\wrt}{with respect to }
\newcommand{\iras}{{\sl IRAS }}
\newcommand{\cobe}{{\sl COBE }}
\newcommand{\Oh}{\myfm{\Omega h}}
%
\newcommand{\etal}{{\em et al.\/ }}
\newcommand{\eg}{{\em e.g.\/ }}
\newcommand{\etc}{{\em etc.\/ }}
\newcommand{\ie}{{\em i.e.\/ }}
\newcommand{\viz}{{\em viz.\/ }}
\newcommand{\cf}{{\em cf.\/ }}
\newcommand{\via}{{\em via\/ }}
\newcommand{\apriori}{{\em a priori\/ }}
\newcommand{\adhoc}{{\em ad hoc\/ }}
\newcommand{\viceversa}{{\em vice versa\/ }}
\newcommand{\versus}{{\em versus\/ }}
\newcommand{\qed}{{\em q.e.d. \/}}
\newcommand{\<}{\thinspace}
%
\newcommand{\km}{\unit{km}}
\newcommand{\kms}{\unit{km~s\mone}}
\newcommand{\kmsa}{\unit{km~s\mone~arcmin}}
\newcommand{\kpc}{\unit{kpc}}
\newcommand{\mpc}{\unit{Mpc}}
\newcommand{\hkpc}{\myfm{h\mone}\kpc}
\newcommand{\hmpc}{\myfm{h\mone}\mpc}
\newcommand{\parsec}{\unit{pc}}
\newcommand{\cm}{\unit{cm}}
\newcommand{\yr}{\unit{yr}}
\newcommand{\au}{\unit{A.U.}}
\newcommand{\AU}{\au}
\newcommand{\gm}{\unit{g}}
\newcommand{\solarm}{\unit{M\sun}}
\newcommand{\Lsun}{\unit{L\sun}}
\newcommand{\Rsun}{\unit{R\sun}}
\newcommand{\seconds}{\unit{s}}
\newcommand{\micro}{\myfm{\mu}}
\newcommand{\Mdot}{\myfm{\dot M}}
%
%
%
\newcommand{\dgr}{\myfm{^\circ} }
\newcommand{\ddgr}{\mbox{\dgr\hskip-0.3em .}}
\newcommand{\mnt}{\mbox{\myfm{'}\hskip-0.3em .}}
\newcommand{\scnd}{\mbox{\myfm{''}\hskip-0.3em .}}
\newcommand{\hr}{\myfm{^{\rm h}}}
\newcommand{\dhr}{\mbox{\hr\hskip-0.3em .}}
%
%
%
%
%
%
%
\newcommand{\refindent}{\par\noindent\hangindent=0.5in\hangafter=1}
\newcommand{\figpar}{\par\noindent\hangindent=0.7in\hangafter=1}
%
%

\newcommand{\mybiblio}{\vspace{1cm}
		       \setcounter{subsection}{0}
		       \addtocounter{section}{1}
		       \def\junksec{References} 
 }

%
%
%

%
%
%
%
%

\newcommand{\vol}[2]{ {\bf#1}, #2}
\newcommand{\jour}[4]{#1. {\it #2\/}, {\bf#3}, #4}
\newcommand{\physrevd}[3]{\jour{#1}{Phys Rev D}{#2}{#3}}
\newcommand{\physrevlett}[3]{\jour{#1}{Phys Rev Lett}{#2}{#3}}
\newcommand{\aaa}[3]{\jour{#1}{A\&A}{#2}{#3}}
\newcommand{\aaarev}[3]{\jour{#1}{A\&A Review}{#2}{#3}}
\newcommand{\aaas}[3]{\jour{#1}{A\&A Supp.}{#2}{#3}}
\newcommand{\aj}[3]{\jour{#1}{AJ}{#2}{#3}}
\newcommand{\apj}[3]{\jour{#1}{ApJ}{#2}{#3}}
\newcommand{\apjl}[3]{\jour{#1}{ApJ Lett.}{#2}{#3}}
\newcommand{\apjs}[3]{\jour{#1}{ApJ Suppl.}{#2}{#3}}
\newcommand{\araa}[3]{\jour{#1}{ARAA}{#2}{#3}}
\newcommand{\mn}[3]{\jour{#1}{MNRAS}{#2}{#3}}
\newcommand{\mnras}{\mn}
\newcommand{\jgeo}[3]{\jour{#1}{Journal of Geophysical Research}{#2}{#3}}
\newcommand{\qjras}[3]{\jour{#1}{QJRAS}{#2}{#3}}
\newcommand{\nat}[3]{\jour{#1}{Nature}{#2}{#3}}
\newcommand{\pasa}[3]{\jour{#1}{PAS Australia}{#2}{#3}}
\newcommand{\pasj}[3]{\jour{#1}{PAS Japan}{#2}{#3}}
\newcommand{\pasp}[3]{\jour{#1}{PAS Pacific}{#2}{#3}}
\newcommand{\rmp}[3]{\jour{#1}{Rev. Mod. Phys.}{#2}{#3}}
\newcommand{\science}[3]{\jour{#1}{Science}{#2}{#3}}
\newcommand{\vistas}[3]{\jour{#1}{Vistas in Astronomy}{#2}{#3}}

%
%
%
\newcommand{\leftb}{<\!\!} \newcommand{\rightb}{\!\!>}
\newcommand{\oversim}[2]{\protect{\mbox{\lower0.5ex\vbox{%
  \baselineskip=0pt\lineskip=0.2ex
  \ialign{$\mathsurround=0pt #1\hfil##\hfil$\crcr#2\crcr\sim\crcr}}}}} 
\newcommand{\simgreat}{\mbox{$\,\mathrel{\mathpalette\oversim>}\,$}} 
\newcommand{\simless} {\mbox{$\,\mathrel{\mathpalette\oversim<}\,$}} 
%
%
%
%
\title{On the origin of brown dwarfs and free-floating planetary 
mass objects}  
\author[Kroupa, Bouvier]{Pavel Kroupa$^{1,2,3}$, Jerome Bouvier$^1$
\\
$^1$Laboratoire d'Astrophysique de l'Observatoire de Grenoble, BP 53,
   F-38041 Grenoble Cedex 9, France \\
$^2$Institut f\"ur Theoretische Physik und Astrophysik der
Universit\"at Kiel, D-24098 Kiel, Germany\\
$^3$ {\it Heisenberg Fellow}
}

\maketitle

\begin{abstract} 
Briceno et al. report a significantly smaller number of brown dwarfs
(BDs) per star in the Taurus-Auriga (TA) pre-main sequence stellar
groups than in the central region of the Orion Nebula cluster (ONC).
Also, BDs have binary properties that are not compatible with a
star-like formation history.  It is shown here that these results can
be understood if BDs are produced as ejected embryos with a dispersion
of ejection velocities of about 2~km/s and if the number of ejected
embryos is about one per four stars born in TA and ONC. The Briceno et
al. observation is thus compatible with a universal BD production
mechanism and a universal IMF, but the required number of BDs per star
is much too small to account for the one BD per star deduced to be
present in the Galactic field.  There are two other mechanisms for
producing BDs and free-floating planetary-mass objects (FFLOPs),
namely the removal of accretion envelopes from low-mass proto-stars
through photo-evaporation through nearby massive stars, and hyperbolic
collisions between proto-stars in dense clusters. The third BD
flavour, the collisional BDs, can be neglected in the ONC.  It is
shown that the observed IMF with a flattening near $0.5\,M_\odot$ can
be re-produced via photo-evaporation of proto-stars if they are
distributed according to a featureless Salpeter MF above the
sub-stellar mass limit, and that the photo-evaporated BDs should have
a smaller velocity dispersion than the stars. The number of
photo-evaporated BDs per star should increase with cluster mass,
peaking in globular clusters that would have contained many stars as
massive as $150\,M_\odot$.  The required number of embryo-ejected BDs
in TA and the ONC can be as low as 6~ejected BDs per 100 stars if the
central ONC contains 0.23~photo-evaporated BDs per
star. Alternatively, if the assumption is discarded that embryo
ejection must operate equally in all environments, then it can be
argued that TA produced about one ejected BD per star leading to
consistency with the Galactic-field observations. The dispersion of
ejection velocities would be about 3~km/s.  In the central ONC the
number of ejected BDs per star would then be at most~0.37, or less if
photo-evaporated BDs contribute. This non-universal scenario would
thus imply that the Galactic-field BD population may mostly stem from
TA-like star formation or modest clusters, the ONC not being able to
contribute more than about~$0.25\pm0.04$ BDs per star.

{\keywords stars: formation --
stars: low-mass, brown dwarfs -- binaries: general
-- open clusters and associations: general -- Galaxy: stellar content}

\end{abstract}

\section{Introduction}
\label{sec:intro}

The failure until a few years ago to find sub-stellar-mass objects
($m<0.08\,M_\odot$) posed a long-standing unsolved problem because
theoretical considerations (e.g. Boss 1986, 2001; Kumar 2001)
suggested that opacity-limited fragmentation may proceed down to
$0.001-0.01\,M_\odot$.  According to this argument, a dense molecular
cloud region contracts with constant temperature and rising density so
that the Jeans mass becomes smaller allowing smaller fragments to form
as time progresses. The collapse leads to an increasing opacity which
ultimately increases sufficiently to prohibit radiative cooling of the
core. The core heats up and the Jeans mass increases, thus leading to
the above minimum fragmentation mass. On the other hand, arguments had
been put forward (e.g. Adams \& Fatuzzo 1996) that the hydrostatic
core, which forms within a fragment once opacity stops significant
radiative cooling, would continue to accrete from an envelope, which
always exceeds the hydrostatic core mass by large factors, until
feedback reverses the infall. Feedback energy comes from collimated
outflows and the luminosity of the protostar due to accretion
luminosity, deuterium burning and finally hydrogen burning.  The
typical conditions in dense molecular cloud regions are such that the
formation of sub-stellar mass objects, that require very feeble
mass-accretion rates, is unlikely. This fitted rather well with the
non-detection of these objects, and one could begin feeling somewhat
comfortable with this null result.

However, since a few years ago sub-stellar-mass objects are being
discovered in increasing numbers (e.g. Basri 2000).  Sub-stellar-mass
objects can be split into two broad categories, brown dwarfs (BDs,
$0.01\simless m/M_\odot \simless 0.08$) and low-mass BDs orbiting a
star or a BD, or free-floating planetary-mass objects (FFLOPs, $m
\simless 0.01\,M_\odot \approx 10\,M_{\rm J}$, $M_{\rm J}=$~Jupiter
mass). Differentiating between massive planets and low-mass BDs
orbiting a star is difficult observationally. Theoretically, they can
be distinguished according to whether their formation involved the
initial condensation of solids in a circum-stellar disk (yielding
planets) or the collapse of a cloud or disk fragment (e.g. Kumar
2002).  The internal structure and constitution of BDs and FFLOPs is
over-viewed by Chabrier \& Baraffe (2000).

BDs and FFLOPs have been found in a number of very young clusters
where they are most easily detected given that they fade with time
(Lucas \& Roche 2000; Zapatero Osorio et al. 2000, Muench, Lada \&
Lada 2000; Bouvier et al. 2002; Moraux et al. 2003; Muench et
al. 2002, 2003).  Measurements of the mass functions (MFs) in five
clusters indicate that their distribution can be described with a
power-law function, $\xi(m)\propto m^{-\alpha}$, where $\xi(m)\,dm$ is
the number of stars (or BDs and FFLOPs) in the mass interval $m,m+dm$,
and $\alpha = +0.5\pm 0.1$ between about~0.03 and $0.3\,M_\odot$
(Bouvier et al. 2002). The initial MF (IMF) of stars can be
represented by the standard form $\alpha=+1.3\pm0.5, 0.08 \le
m/M_\odot \le 0.5$ and $\alpha=+2.3\pm0.3, 0.5 < m/M_\odot$ (Kroupa
2002). Such structure in the IMF contains information on the processes
active in the conversion of interstellar gas to stars, BDs and
FFLOPs. And in particular, variations of structure in the IMF between
different star-formating regions, if found, would place important
constraints on our understanding of how their formation proceeds.

The discovery by Briceno et al. (2002) that star formation in
Taurus--Auriga (TA) seems to be producing significantly fewer BDs,
with masses in the range $0.02-0.08\,M_\odot$, per low-mass stellar
system than the star-formation event that formed the much denser Orion
Nebula cluster (ONC) may thus pose a real break-through. According to
this result, the IMF for BDs (and probably FFLOPs) may be dependent on
environment, while the stellar IMF seems to be invariant and
consistent with the standard form (fig.~11 in Briceno et al. 2002;
Kroupa et al. 2003, hereinafter KBDM).

The reported variation in the BD mass regime may be a result of
star-formation proceeding under different conditions. For example, the
Jeans mass was much smaller in the ONC precursor than in TA which is
at least qualitatively consistent with the smaller number of BDs per
star seen in TA (Briceno et al. 2002). Indeed, a number of authors
have found evidence that BDs form just like stars.  The measured
longevity of circum-BD disks is generally taken to imply disk masses
that are not consistent with having been truncated through ejections
(Liu et al. 2002), while White \& Basri (2002) discard the ejection
hypothesis because they also find young BDs to posses accretion disks
and the BDs in TA to have indistinguishable kinematics from the
stars. Based on the measured properties of the dust disk around the BD
GY11 in $\rho$~Oph, Testi et al. (2002) conclude that BDs form like
stars by core contraction, gravitational collapse and accretion, while
most BD candidates in the central region of the ONC and in $\rho$~Oph
also appear to sport disks thus likewise indicating a similar origin
as stars (Muench et al. 2001; Natta et al. 2002).

If BDs have the same general formation history as stars then they will
have formed from very-low mass cloud cores and ought to have binary
properties that are a natural extension of those of late-type stars
(Delgado-Donate, Clarke \& Bate 2003). If this were to be the case
then a possible reason as to why the number of BDs is smaller in TA
than in ONC may be because the ONC is dynamically more evolved. In TA
BDs may remain locked-up in binary systems.  Indeed, KBDM show that
the observed different number of BDs per stellar system in TA and the
ONC is obtained nearly exactly if it is assumed that BDs and stars
form according to the same rules, i.e. that the initial pairing
properties of BDs and stars do not differ. These pairing properties
are described by the {\it standard model of star formation} extended
into the sub-stellar mass range.  According to this model stars and
BDs form in binaries with companion masses picked randomly from the
IMF, i.e. that there is no sudden change in pairing properties below
the hydrogen-burning mass limit.  Encounters in dense clusters (such
as the ONC) preferentially destroy binaries with weaker binding energy
leading to a larger number of free-floating BDs than in environments
that are dynamically unevolved (such as TA). This process has also
been studied by Adams et al. (2002) for the Pleiades and Hyades
clusters.

While the standard model with BDs arrives at the correct relative
number of BDs per stellar system, it also makes predictions on the
binary fraction of stellar and BD systems and their distribution of
semi-major axes.  Comparison of these predictions with the available
binary statistics demonstrates that the standard model with BDs cannot
be the solution to the differences seen by Briceno et
al. (2002). Changing the IMF (as suggested by Briceno et al.) to model
its possible dependence on the Jeans mass does not lead to a model
which is consistent with both the number of BDs per stellar system and
with the binary statistics in the TA star-formation rate.

The conclusion is therefore that in TA the BDs follow different
pairing rules, i.e. that they cannot have formed with the same binary
properties as stars. The standard model with BDs fails.  Because the
standard model (without BDs) leads to excellent agreement to the
observational data in TA, the ONC and Pleiades, as well as the
Galactic-field stellar population, KBDM suggest {\it that BDs need to
be added as a separate, primarily single-object population}. This is
supported by the conclusions made by Close et al. (2003) based on
their survey of the binary properties of very-low mass stars and BDs
in the solar neighbourhood.  Thus, BDs do not appear to have the same
general formation history as stars.

There exist four broad formation scenarios for BDs and FFLOPs: (i) in
contracting cloud fragments like stars with accretion from an envelope
until feedback from the hydrostatic core halts accretion (the {\it
star-like model}), (ii) through ejection from dynamically unstable
multiple-proto-stellar systems and the consequent loss of the
accretion envelope (the {\it embryo-ejection model}), (iii) removal of
the accretion envelope due to photo-evaporation through a nearby
massive star (the {\it photo-evaporation model}), or (iv) separation
of embryos from their accretion envelopes through hyperbolic
encounters in dense embedded clusters (the {\it collision model}).
Note that according to scenarios (ii)--(iv) the star-like formation
process (i) is terminated through the loss of the accretion
envelope. Another scenario for the formation of FFLOPs as being
planets that were lost from their parent stars through
stellar-dynamical encounters cannot operate to produce a significant
number of the observed FFLOPs in young clusters, because the
cross-section of typical planetary systems is too small and the
potential wells of the clusters are too shallow to retain the
so-produced FFLOPs (Smith \& Bonnell 2001).

Only scenarios (i) and (ii) can be valid in the TA star-formation rate
because there are no hot ionising stars there and the densities are
too low, but (i) is unlikely to be a major BD production channel given
the results of KBDM: There are virtually no BD binaries in TA, but a
high BD binary fraction would be expected.  In the ONC all four
scenarios may have played a role (binary statistics do not yet exist
for the BDs and FFLOPs in the ONC). However, it would appear as very
unlikely that scenario (i) may have acted in the ONC but not in the
much more tranquil TA star-formation region which is more likely to
allow the subtle conditions leading to the very low accretion rates
needed to assemble a BD system.  Scenario (i) can therefore probably
be excluded as a significant source of sub-stellar mass objects. This
conclusion is consistent with the finding by Close et al. (2003),
Gizis et al. (2003), Bouy et al. (2003) and Mart{\'{\i}}n et
al. (2003) that very-low-mass solar-neighbourhood and Pleiades stars
and BDs appear to have binary properties that are inconsistent with an
extrapolation of those of M--G~dwarfs, or with a scaling to T~Tauri
stars.

In what follows the embryo-ejection scenario and its implications are
considered (\S~\ref{sec:ej}), followed by an investigation of the
implications of the photo-evaporation model (\S~\ref{sec:phot}).
Section~\ref{sec:rem} discusses the collision model.  A combination of
the three models is studied in \S~\ref{sec:comb}, and \S~\ref{sec:fin}
finds that the birth-rate per star of BDs may differ significantly in
different star-forming environments if the Galactic-field BD density
has been correctly estimated.  The conclusions are presented in
\S~\ref{sec:concs}.

\section{The embryo-ejection model}
\label{sec:ej}

If massive proto-stellar disks with radii $\approx 100$~AU fragment
rapidly to form many massive planets then this system relaxes within
100~orbits ($\approx 10^5$~yr) by ejecting most of the planetary
siblings that become FFLOPs into the cluster leaving one to a few
bound to the parent star (Papaloizou \& Terquem 2001).  The remaining
star-planet system will typically have the most massive planet on an
eccentric, short-period orbit. The dramatic rise of the BD MF below
$0.03\,M_\odot$ found by Muench et al. (2002, 2003) in the central ONC
and in IC~348 may pose tentative support for this scenario.

According to the recently emphasised embryo-ejection hypothesis
(Reipurth \& Clarke 2001), BDs (and some FFLOPs) are unfinished stars
that were ejected from their natal embedded system (Reipurth 2000;
Bate, Bonnell \& Bromm 2002; Delgado-Donate, Clarke \& Bate 2003). The
fragmentation of a cloud core (or kernel according to Myers 1998) with
a mass of $M=1\,M_\odot$ and a radius of $R=1000$~AU typically forms a
few accreting hydrostatic cores.  A non-hierarchical system of a few
bodies is dynamically unstable and decays within a few system
dynamical times,
\begin{equation}
t_{\rm dyn}\approx 0.5\,M_{[M_\odot]}^{-1/2}\,R_{\rm [AU]}^{3/2} =
1.6\times 10^4 \quad {\rm yr},
\label{eq:tdyn}
\end{equation} 
by typically ejecting the least-massive member (Sterzik \& Durisen
1998). Ejections continue on the new shorter dynamical time-scale of
the shrunk (hardened) system leaving one binary or a strongly
hierarchical and thus long-lived multiple system. 

The one high-resolution computation of a fragmenting cloud by Bate,
Bonnell \& Bromm (2002, 2003) confirms that the ejection of unfinished
stars occurs rather often.  This computation models the fragmentation
of a small proto-cluster, and is thus applicable to star formation in
TA rather than in the ONC.  The computation suggests that about the
same number of BDs are formed as low-mass stars.  This large
production rate of about one~BD per star may decrease in a computation
that is allowed to proceed for a longer time, in which case some of
the hydrostatic cores may accrete sufficient mass to become
stars. Furthermore, the presently feasible hydrodynamical collapse
computations lack feedback through radiation, winds and outflows,
which in reality begin to heat a collapsing cloud as soon as the first
accreting hydrostatic core forms, thus limiting the number density of
accreting hydrostatic cores, and therefore the ejection rate of
unfinished stars may be reduced.  {  On the other hand, radiation
and outflows may trigger new formation or destroy envelopes.}  It will
be a very serious challenge to include these computationally extremely
costly effects in future collapse calculations.  {\it The real
production rate of BDs as ejected embryos may therefore differ from
what the presently available hydrodynamical computations suggest}.

The predictions of the embryo-ejection model are as follows: (i) The
typical ejection velocities are $v_{\rm ej} \approx 15\,R_{\rm
C,[AU]}^{-1/2} \simless 2$~km/s for closest-approach distances $R_{\rm
C}\simgreat 56$~AU, but a high-velocity tail exists (Sterzik \&
Durisen 1995; Reipurth \& Clarke 2001; Sterzik \& Durisen 2003;
Delgado-Donate et al. 2003).  (ii) The so-formed BDs have a very small
binary fraction, although BD--BD binaries with semi-major axis $a
\simless 0.5\times R_{\rm C}$ can survive the ejection process, and
(iii) the ejected BDs can also retain accretion disks with radii
$<0.5\times R_{\rm C}$. The hydrodynamical collapse calculations
reported by Bate et al. (2002; 2003) show that most (14) of their 18
BDs are ejected and have disks with radii $\simless 10$~AU.  The low
binary proportion is qualitatively consistent with available surveys
for BDs, and while a large fraction of young BDs have infrared
excesses indicative of disks (Muench et al. 2001; Liu, Najita \&
Tokunaga 2002; Natta et al. 2002), their sizes are not yet known.

If embryo ejection is the dominant source of most BDs in TA and in the
ONC, then we would like to know if the different gravitational
potentials of the star-forming regions (TA vs ONC) can lead to the
observed difference between the fraction of BDs that are retained in
the stellar group or star cluster (Bouvier et al. 2001). This is
estimated here by making the ansatz that embryo ejection occurs
exactly alike in the vastly different TA and ONC environments by
requiring the number of BDs per star to be the same and that the
ejected embryos have, as a population, an isotropic Schwarzschild
velocity distribution function with a one-dimensional velocity
dispersion $\sigma_{\rm 1D}$,
\begin{equation}
h(v) = {1\over (2\,\pi\,\sigma_{\rm 1D}^2)^{3/2}}\,e^{-{1\over 2}
       \left( {v\over \sigma_{\rm 1D}}  \right)^2},
\label{eq:schw}
\end{equation}
where $v$ is the 3D speed.  The ansatz is basically the assumption
that embryo ejection occurs from fragmented cloud kernels that have
the same statistical properties on scales below a few hundred~AU in TA
and the proto-ONC.  {  Both the TA groups and the ONC are about
1~Myr old so that the vast majority of dynamical decays will have
occurred to completion and therefore very few if any further BDs are
expected to be produced.}  The velocity dispersion is made up (i) of
contributions from kernel--kernel motions, i.e. the velocity
dispersion of the embedded cluster with mass $M_{\rm ecl+g}$ (stars
plus gas) which is approximately in global virial equilibrium (Kroupa
\& Boily 2002), and (ii) the ejected BDs,
\begin{equation}
\sigma_{\rm 1D}^2 = \sigma_{\rm cl,1D}^2 + \sigma_{\rm ej,1D}^2,
\label{eq:sigma}
\end{equation}
where $\sigma_{\rm cl,1D}^2=G\, M_{\rm ecl+g} / 2\, R_{0.5}$,
$G=0.0045\,$pc$^3/(M_\odot\,$Myr$^2)$ is the gravitational constant
and $R_{0.5}$ is the half-mass radius.  According to the above ansatz
$\sigma_{\rm ej,1D}$ is to be equal in both environments. 

A star-formation event produces a total number of BD systems, $N_{\rm
BD,tot}$. { According to the results of KBDM we assume these to have
the same binary properties independent of environment and the number
of BD companions to stars to be negligible. Most BDs are single, with
only about 15~per cent or fewer being binaries with semi-major axes
$<15$~AU (Close et al. 2003; Gizis et al. 2003; Bouy et al. 2003;
Mart{\'{\i}}n et al. 2003). In what follows we do not distinguish
between BD binaries and single BDs, although we note that on average
binary BDs should have lower ejection speeds than single BDs by virtue
of their higher system masses and necessary lower ejection speeds to
guarantee binary survival.  Of the total number of BDs, $N_{\rm
BD,obs}$ are observed. These can be split into two populations: The
bound BDs with velocities smaller than the escape speed, $v_{\rm
esc}$, from the stellar group or cluster, and the unbound BDs which
remain within the survey region.  The number of unbound but observable
BDs can be estimated by calculating the number of BDs with velocities
in the interval $v_{\rm esc}$ to $v_{\rm in}$, where $v_{\rm in}$ is
given roughly by $v_{\rm in}=r_{\rm in}/\tau$. Here $r_{\rm in}$ is
the radius of the survey volume ($r_{\rm in,TA}=2.3$~pc for the TA
groups, while $r_{\rm in,ONC}=0.5$~pc for the central ONC), and the
age of both the TA and ONC is about $\tau=1$~Myr.}  Thus,
\begin{eqnarray}
N_{\rm BD,obs} & = &N_{\rm BD,tot}\,\left(
\int_0^{v_{\rm esc}} h(v)\,4\,\pi\,v^2\,dv + 
\int_{v_{\rm esc}}^{v_{\rm in}} h(v)\,4\,\pi\,v^2\,dv \right),\\
 & \equiv & N_{\rm BD,tot}\,\left( {\cal B} + {\cal U} \right)
\label{eq:BDobs}
\end{eqnarray}
where $N_{\rm BD,tot}\,h(v)\,4\,\pi\,v^2\,dv$ is the number of BDs
with speeds in the interval $v$ to $v+dv$. The ``retention integral''
${\cal B} = {\cal B}(\sigma_{\rm ej,1D},M_{\rm ecl+g})$ ($={\cal B}_{\rm
TA}$ for the TA groups or ${\cal B}_{\rm ONC}$ for the ONC), and
similarly the ``unbound but observable integral'' ${\cal U} = {\cal
U}(\sigma_{\rm ej,1D},M_{\rm ecl+g})$ which is zero if $v_{\rm esc} \ge
v_{\rm in}$.

The observational datum is the number of BDs per stellar system in the
group or cluster,
\begin{equation}
{\cal R}_{\rm obs} \equiv {N_{\rm BD,obs} \over N_{\rm st,obs}}.
\label{eq:Robs}
\end{equation}
According to our ansatz each star-formation event produces the same
number of BDs per star,
\begin{equation}
{\cal R} \equiv {N_{\rm BD,tot} \over N_{\rm st,tot}} = {\rm constant},
\label{eq:R2}
\end{equation}
the BDs and stars having masses in some interval as defined by the
observational survey (eq.~\ref{eq:Rimf} below). The number of observed
stars is the number of stellar systems, $N_{\rm st,obs}\equiv N_{\rm
st,sys} = N_{\rm st,bin}+N_{\rm st,sing}$, where the total number of
stars is $N_{\rm st,tot} = 2\,N_{\rm st,bin}+N_{\rm
st,sing}=(1+f)N_{\rm st,sys}$.  The binary fraction in TA is $f_{\rm
TA}=1$ while in the ONC it is $f_{\rm ONC}=0.5$ (approximately).
Thus,
\begin{eqnarray}
{\cal R}_{\rm obs} 
     & = & {N_{\rm BD,tot}\over N_{\rm st,sys}}
           \, \left( {\cal B} + {\cal U} \right), \nonumber\\
     & = & (1+f) \, {\cal R} \,\left( {\cal B} + {\cal U} \right) .
\label{eq:R3}
\end{eqnarray}
With ${\cal R}_{\rm obs, TA}=0.17\pm0.06$ for TA and ${\cal R}_{\rm
obs, ONC}=0.38\pm0.06$ for the ONC (KBDM, from the data provided by
Briceno et al. 2002) the ratio ${\cal R}_{\rm obs, TA}\,(1+f_{\rm
TA})^{-1}/ {\cal R}_{\rm obs, ONC}\,(1+f_{\rm ONC})^{-1} \equiv
\Lambda_{\rm obs}=0.34\pm0.05$, where
\begin{equation}
\Lambda(\sigma_{\rm ej,1D}) \equiv  
              {{\cal R}\,\left({\cal B}_{\rm TA} + {\cal U}_{\rm TA}\right) 
              \over 
               {\cal R}\,\left({\cal B}_{\rm ONC} + {\cal U}_{\rm ONC}\right)}.
\label{eq:cond}
\end{equation}
The escape speed from the centre of a Plummer sphere, which
approximates the structure of the embedded groups and cluster
reasonably well for the present purpose, is
\begin{equation}
v_{\rm esc} = 1.62\, \left( {G\,M_{\rm ecl+g} \over R_{0.5}}
                     \right)^{1\over2}.
\label{eq:vesc}
\end{equation}
Assuming a star-formation efficiency of 33~per cent, the mass (stars
plus gas) of the typical embedded TA group is approximately $M_{\rm
ecl+g,TA}=50\,M_\odot$ (Kroupa \& Bouvier 2003) while the embedded
cluster mass of the ONC precursor was about $M_{\rm
ecl+g,ONC}=9000\,M_\odot$ (Kroupa, Aarseth \& Hurley 2001, hereinafter
KAH). The half-mass radii of the embedded groups and cluster are
approximately $R_{\rm 0.5,TA} = 0.3$~pc and $R_{\rm 0.5,ONC} =
0.4$~pc, giving $v_{\rm esc,TA} = 1.4$ and $v_{\rm esc,ONC} =
16.3$~km/s, respectively.  {  The embedded ONC model used here
remains embedded for 0.6~Myr. Then the gas is removed on a thermal
time-scale, and the cluster evolves to the observed ONC at an age of
about 1~Myr and to the observed Pleiades at an age of 100~Myr (KAH).}
Once the gas is removed the ONC expands, but the BDs retained during
the embedded phase with $v<v_{\rm esc}$ will behave like the stars so
that $R_{\rm obs}$ is conserved approximately during the
expansion. {  This will hold true if most of the BDs are ejected
early such that the bound BD population has time to relax with the
stars, and provided the velocity dispersion of ejected BDs is smaller
than the velocity dispersion of stars in the virialised but embedded
cluster, $\sigma_{\rm ej,1D} < \sigma_{\rm cl,1D} \approx 5$~km/s
(KAH), so that the velocity distribution function of BDs and stars
will not differ much. The decay of the small$-N$ groups occurs on a
time-scale of a few$\times10^4$~yr $\ll 1$~Myr so that the production
of most BDs occurs well before the end of the embedded phase, and we
will see that $\sigma_{\rm ej,1D} < \sigma_{\rm cl,1D}$ is indeed the
case. The TA groups do not evolve significantly during their first~Myr
(Kroupa \& Bouvier 2003).}

The aim is now to seek the velocity dispersion $\sigma_{\rm ej,1D}$
which minimises $|\delta|$, where
\begin{equation}
\delta \equiv \Lambda_{\rm obs}-\Lambda(\sigma_{\rm ej,1D}).
\label{eq:delta}
\end{equation}
To this end, the retention and unbound integrals in eq.~\ref{eq:BDobs}
are solved numerically.
\begin{figure}
\begin{center}
\rotatebox{0}{\resizebox{0.6 \textwidth}{!}
{\includegraphics{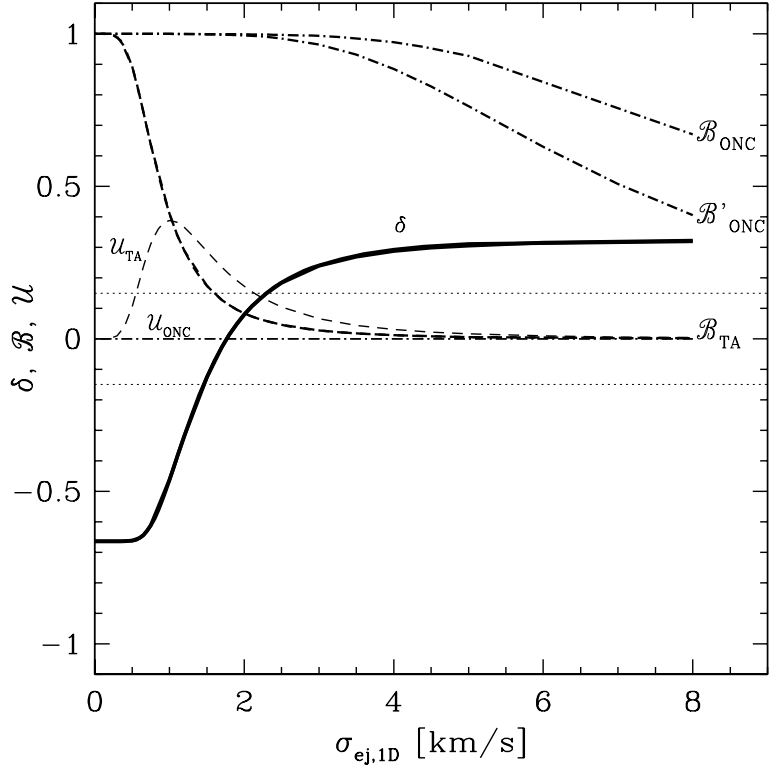}}}
\vskip -20mm
\caption
{The dependence of the bound fraction of BDs, ${\cal B}$, and of the
unbound but observable fraction, ${\cal U}$, on the one-dimensional
dispersion of ejection velocities is shown for a typical TA group as
the short-dashed curve, while for the embedded pre-ONC model it is
shown as dot-dashed lines (${\cal B}_{\rm ONC}$ is for $M_{\rm
ecl+g}=9000\,M_\odot$ while ${\cal B}^{'}_{\rm ONC}$ is for $M_{\rm
ecl+g}=4500\,M_\odot$). The difference $\delta$ (eq.~\ref{eq:delta}) is
plotted as the thick solid curve, while the thin dotted lines indicate
the observational three-sigma range on $\delta$. Note that $\delta$
does not depend sensitively on the mass of the ONC.}
\label{fig:fr}
\end{center}
\end{figure} 

The results are displayed in Fig.~\ref{fig:fr}. More than 80~per cent
of all BDs are lost from the typical TA group if $\sigma_{\rm
ej,1D}\simgreat 1.5$~km/s, while the embedded pre-ONC retains most of
its BDs for $\sigma_{\rm ej,1D}\simless 6$~km/s by virtue of its much
deeper potential well. The truly interesting result, however, is that
there exists one ejection velocity, $\sigma_{\rm
ej,1D}=2.0^{+0.3}_{-0.5}$~km/s (three-sigma interval: $\delta \in
(-0.15,0.15)$), which simultaneously leads to the observed number of
BDs per star in the TA groups and the ONC. Near this velocity there
are about two times more unbound BDs in the survey areas of Briceno et
al. (2002) than BDs bound to the groups, ${\cal U}_{\rm TA}\approx
2\times {\cal B}_{\rm TA}$, while the central region of the ONC only
contains BDs bound to the cluster, ${\cal U}_{\rm ONC}=0.0, {\cal
B}_{\rm ONC} = 1.0$.  Changing the mass of the pre-ONC object to
$M_{\rm ecl+g,ONC}=4500\,M_\odot$ changes the escape speed to $v_{\rm
esc,ONC}=11.5$~km/s but has otherwise a negligible effect on the
result (e.g. $\delta = 0.288$ instead of 0.292 for $\sigma_{\rm ej,1D}
= 4.0$~km/s, {  the difference is not evident in Fig.~\ref{fig:fr}}).

The number of BDs produced per formed (${\cal R}$) star can now be
estimated.  From eq.~\ref{eq:R3} we obtain
\begin{eqnarray}
0.17\pm0.06 &= 2\,{\cal R}_{\rm TA}\,(0.10+0.20) &\Longrightarrow 
{\cal R}_{\rm TA}=0.28\pm0.10, \nonumber\\
0.38\pm0.06 &= 1.5\,{\cal R}_{\rm ONC}\,(1.0+0.0) &\Longrightarrow
{\cal R}_{\rm ONC}=0.25\pm0.04,
\label{eq:Rres}
\end{eqnarray}
so that ${\cal R}_{\rm TA}={\cal R}_{\rm ONC}={\cal R}=0.21-0.29$.
This means that the formation of three--five stars leads, on average,
to the formation of one BD. The mental image corresponding to this
estimate is that {\it a cloud kernel fragments into a number of
hydrostatic cores, which on average leads to the formation of two
binary stellar systems that are weakly bound to each other, and one
ejected unfinished embryo}. 

{  Returning to eq.~\ref{eq:BDobs} we note that by adopting $r_{\rm
in}=2.3$~pc instead of 1.15~pc for the radius of the
2.3~pc~$\times$~2.3~pc TA survey area the contribution of unbound but
visible BDs (${\cal U}_{\rm TA}$) is overestimated. This overestimate
is not serious however, because the model constructed here assumes
that the BDs are ejected at $t=0$ which in reality will not be the
case. Some of the BDs ejected during late times will remain within the
TA survey areas even if the ejection velocity is larger than the
formal $v_{\rm in}$. As an extreme example of the relative
insensitivity of the results, $r_{\rm in}=0$ (no unbound BDs
whatsoever in the survey area, ${\cal U}_{\rm TA}=0$) leads to similar
results: $\sigma_{\rm ej,1D}\approx1.3$~km/s and ${\cal R}_{\rm
TA}\approx{\cal R}_{\rm ONC}\approx 0.25$.}

The calculated BD production rate (per star) can be used to estimate
the sub-stellar IMF. Writing the IMF as a three-part power law (Kroupa
2002),
\begin{equation}
\xi (m) = k\left\{
          \begin{array}{l@{\quad\quad,\quad}l}
   \left({m\over 0.08}\right)^{-\alpha_0}  &m/M_\odot \le 0.08, \\
   \left({m\over 0.08}\right)^{-1.3}  &0.08 < m/M_\odot \le 0.5, \\
   \left[\left({0.5\over 0.08}\right)^{-1.3}\right]
        \left({m\over 0.5}\right)^{-2.3}
        &0.5 < m/M_\odot,\\
          \end{array}\right.
\label{eq:imf_mult}
\end{equation}
where $k$ contains the desired scaling, it follows that
\begin{equation}
{\cal R} = {\int_{0.02}^{0.08}\xi(m)\,dm \over \int_{0.15}^{1.0}\xi(m)\,dm}.
\label{eq:Rimf}
\end{equation}
The integration limits in eq.~\ref{eq:Rimf} are given by the
observational completeness limits of Briceno et al. (2002).  Thus
\begin{equation}
{\cal R} = 0.21(0.29) \Rightarrow 0.0185(0.0255) = 0.08^{\alpha_0} \,
\left({0.08^{1-\alpha_0} - 0.02^{1-\alpha_0} \over 1-\alpha_0}\right),
\end{equation}
being the case for $\alpha_0=-3.3 ({\cal R}=0.21)$ and $\alpha_0=-2.1
({\cal R}=0.29)$. This is much smaller than the estimate
$\alpha_0=+0.3\pm0.7$ for Galactic-field BDs implying ${\cal R}=0.81$
(\S~\ref{sec:fin}). Such a large observed value for $\alpha_0$ would
need larger ejection speeds of the embryos in order to reduce the
observed number of BDs per star in TA. But by the above analysis we
would find too many BDs per star in the ONC. We shall return to this
in \S~\ref{sec:fin}.

In conclusion, this section introduced an analysis of the number of
BDs per star observed in TA and ONC by assuming that the production
mechanism of BDs is via embryo ejection and is the same in all
environments.  This assumption leads to agreement with the BD data in
TA and the ONC and implies that the observed different number of BDs
per star in the TA groups and in the ONC can be explained (i) if the
production rate of BDs per star in both environments is about one BD
per four stars born, and (ii) if the intrinsic velocity dispersion of
the BD population is the same in both environments, $\sigma_{\rm
ej,1D}\approx2$~km/s. The prediction is that the TA groups ought to be
surrounded by single BDs that have not yet been found because the deep
surveys (Briceno et al. 2002) concentrate on the known stellar
aggregates. The implied IMF for BDs falls off much too steeply when
compared to the BD MF in the Galactic field, unless the empirical
Galactic-field BD density has been overestimated (\S~\ref{sec:fin}).

{  As stated at the end of \S~\ref{sec:intro} three other mechanisms
may add BDs. These are formation from very-low-mass cloud kernels with
a star-like accretion history, through destruction of proto-stellar
accretion envelopes due to photo-evaporation from a nearby massive
star, or through hyperbolic collisions between accreting proto-stars
in dense environments. In the next two sections we consider the latter
two in turn, having already excluded the former as a major source of
BDs.  Photo-evaporation and collisions can only have been active in
the ONC but not in the TA groups. It may therefore be possible that
the observational finding $R_{\rm obs,TA} < R_{\rm obs,ONC}$ may be at
least partially explained by having an additional source of BDs in the
ONC. This would, however, imply that the production rate (per star) of
BDs will be a function of the physical environment, contrary to the
assumption posed above.  The implications of this hypothesis on the
global production rate of BDs will be addressed in
\S~\ref{sec:fin}. Finally we note that whatever the formation
mechanism for BDs is, the ONC cannot have produced more or less than
${\cal R}_{\rm ONC}=0.25\pm0.04$ BDs per star in total because ${\cal
B}_{\rm ONC}=1$ for all reasonable $\sigma_{\rm ej,1D}$.}

\section{The photo-evaporation model}
\label{sec:phot}

{  The evaporation of circum-stellar disks through radiation from nearby
O and B stars has been studied in much detail (Johnstone, Hollenbach
\& Bally 1998; St\"orzer \& Hollenbach 1999; Scally \& Clarke 2001)
since the detection of such a process in the ONC with the HST (O'Dell,
Wen \& Hu 1993).}  Henney \& O'Dell (1999) measure the mass-loss rates
of four ONC stars with envelopes, and find that they could not have
been exposed to the UV flux from the central O6~star
($\theta^1$~Ori~C) for longer than about $10^4$~yr. These particular
objects could be crossing the inner cluster region after spending more
time at larger radii. However, the large proportion of stars with
circum-stellar material does indicate that the destructive irradiation
may indeed have turned on recently, probably through the very recent
emergence or birth of the main ionising star $\theta^1$~Ori~C (Kroupa,
Petr \& McCaughrean 1999).  Matsuyama, Johnstone \& Hartmann (2003)
also estimate a short photo-evaporation time-scale of the observed
circum-stellar material through UV flux from external O~stars in the
central region of the ONC.

The reported short photo-evaporation rates of circum-stellar material
thus suggests that very-low mass proto-stars may be severely affected
if present close to an O~star. One way to produce BDs and FFLOPs may
be through removal of the accretion envelopes from low-mass
hydrostatic cores that otherwise would become very-low-mass stars, by
heating of the outer envelopes by the intense radiation from nearby
O~stars (Kroupa 2001; Whitworth \& Zinnecker 2003; Preibisch, Stanke
\& Zinnecker 2003). {  This process for generating BDs and FFLOPs
can only occur during the first 0.1~Myr of a proto-star's life while
most of the mass of a proto-star is in an accretion envelope,} and can
only occur in rich clusters hosting O~stars and cannot give rise to
the BD population detected in TA.  The large fraction of young BDs
with infrared excesses indicative of disks in the central ONC (Muench
et al. 2001) would then constitute nearly naked embryos, the envelopes
of which have not yet been completely removed.

{  To estimate the effect photo-evaporation of accretion envelopes may
have on the MF we device a very simple model, rather than applying the
elaborate radiation transfer treatment which requires the introduction
of a number of parameters (mass profile and extend, mass flow geometry
and rates, opacities) so as to make the problem tractable and which is
dealt with in detail by Johnstone et al. (1998) and St\"orzer \&
Hollenbach (1999), among others. It will be seen that our estimates
turn out to be reasonable.}

The binding energy of the envelope is compared with the radiation
energy received by a protostar which may have already formed a
hydrostatic core.  The initial binding energy of the protostar with
mass $m_{\rm ps}$ and radius $R_{\rm ps}$ is
\begin{equation}
E_{\rm b,i} = - {G\,m_{\rm ps,i}^2 \over R_{\rm ps}}.
\label{eq:Ebin}
\end{equation}
If the protostar receives an amount of external energy $E_{\rm inp}
\equiv \delta E_{\rm b}$ this may lead to the loss of mass in the
outermost envelope, $\delta m \equiv m_{\rm ps,i} - m_{\rm ps,f} \ge
0$. Writing
\begin{eqnarray}
\delta E_{\rm b} &=& E_{\rm b,i} - E_{\rm b,f} \nonumber\\ 
                   &=& - {G \over
                   R_{\rm ps}}\, \left(m_{\rm ps,i}^2 - m_{\rm
                   ps,f}^2\right),
\label{eq:deltaEbin}
\end{eqnarray}
leads to 
\begin{equation}
\delta m = \left[ 1 - \left( 1- {E_{\rm inp} \over |E_{\rm b,i}|}
                     \right)^{1\over 2} \right] \, m_{\rm ps,i}.
\label{eq:deltaM}
\end{equation}
The incident energy received by the protostar over a time-span 
$\delta t$ is
\begin{equation}
E_{\rm inp} = \kappa \, \, \phi\,\, L_{\rm u} \, \delta t,
\label{eq:Einp}
\end{equation}
where the primary ionising star has a luminosity $L_{\rm u}$, $\kappa$
is an efficiency used here to take into account that only a small
fraction of emitted photons will be converted to heat energy in the
proto-stellar envelope. If the protostar is situated a distance $R$
from the O~star only a fraction
\begin{equation}
\phi = {\pi\, R_{\rm ps}^2 \over 4\,\pi\,R^2}
\label{eq:phi}
\end{equation}
of the emitted energy is received by the protostar.  For example, an
O5 star ($L_{\rm u}=10^{5.4}$~erg/s) inputs sufficient energy over
$\delta t=0.1$~Myr into the envelope of a protostar with $R_{\rm
ps}=10$~AU and mass $m_{\rm ps}=0.3\,M_\odot$ on a circular orbit
around the O star with radius $R=0.3$~pc to remove 0.6~per cent of the
proto-star's mass, assuming $\kappa=0.1$.  {  This is essentially the
same result as obtained from a much more sophisticated treatment of
photo-evaporation: From eqn.~2 in Scally \& Clarke (2001), which is
applicable for an external radiation field dominated by far
ultraviolet photons, a mass loss of 0.6~per cent is arrived at over
0.1~Myr}.

\begin{figure}
\begin{center}
\rotatebox{0}{\resizebox{0.6 \textwidth}{!}
{\includegraphics{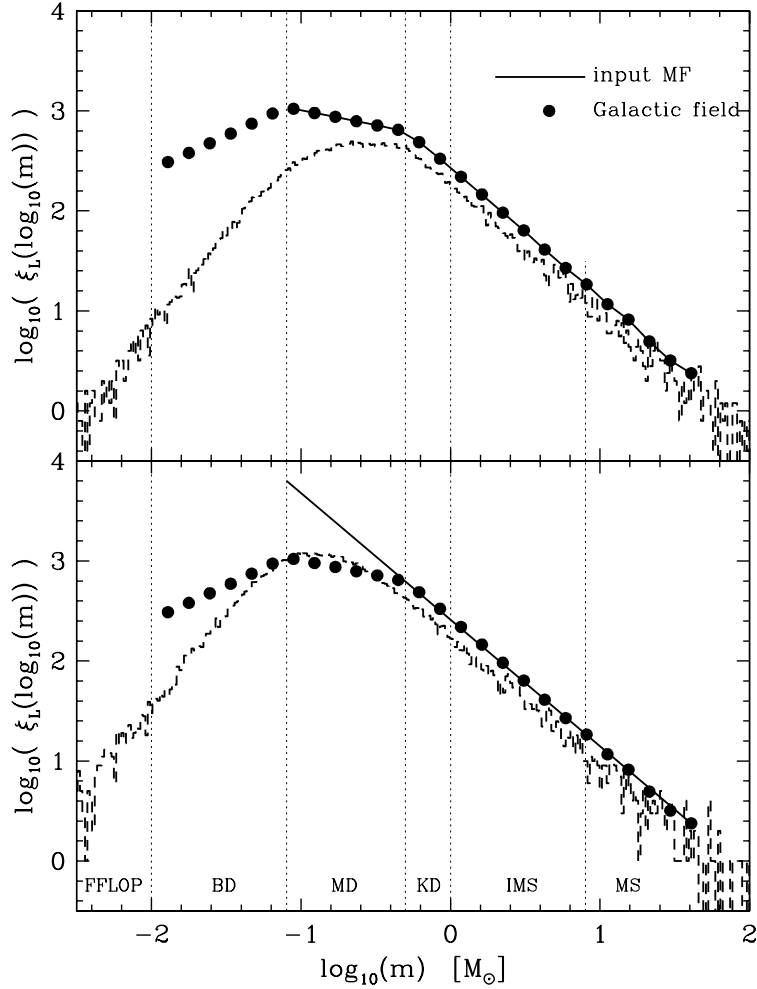}}}
\vskip 0mm
\caption
{The IMF resulting from the photo-evaporation envelope-destruction
model. In the upper panel the input proto-stellar MF is the standard
IMF with $m\ge 0.08\,M_\odot$, while in the lower panel a Salpeter
proto-stellar MF for $m\ge 0.08\,M_\odot$ is adopted.  In both panels
the dots are the standard or average Galactic-field IMF for $m\ge
0.01\,M_\odot$. The model assumes $\delta t=0.1$~Myr, $\kappa=0.1$,
$R=0.1$~pc, $L_{\rm u}=10^{5.4}\,L_\odot$ (an O5 star). The vertical
dotted lines delineate simplified object types (MD,KD $=$ M,K~dwarf,
IMS $=$ intermediate mass star, MS $=$ massive star). }
\label{fig:mf}
\end{center}
\end{figure}

The overall effect for an ensemble of $10^5$~proto-stars on circular
orbits with $R=0.1$~pc is illustrated in Fig.~\ref{fig:mf} for two
proto-stellar MFs {  that are truncated at the hydrogen-burning mass
limit. Such a truncation is suggested by the reasoning of
\S~\ref{sec:intro} that BDs do not appear to have the same formation
history as stars and therefore appear to originate through a mechanism
that differs from that of most stars which presumably manage to
accrete their available gas reservoir.}  The figure shows that if the
proto-stellar MF is identical to the standard Galactic-field stellar
IMF, then after 0.1~Myr the resulting stellar IMF may contain, within
0.1~pc of the main ionising star, far too few M~dwarfs and BDs
compared to observational constraints on the IMF in the central ONC
for example (Muench et al. 2001). If, however, the proto-stellar MF is
a Salpeter MF ($\alpha=+2.3, m\ge 0.08\,M_\odot$), then the observed
stellar IMF may be obtained, but a deficit of BDs relative to the
Galactic-field IMF may nevertheless be evident. Note that the results
arrived at here may over-estimate photo-evaporation for $m_{\rm
ps,i}\simgreat 0.3\,M_\odot$ relative to less-massive proto-stars
because their shorter dynamical time and thus shorter collapse-phase
has not been taken into account, in order not to introduce additional
parameters into the model. Instead, it was assumed that proto-stars
spend 0.1~Myr in an extended state independent of $m_{\rm ps,i}$.

A single pre-stellar power-law MF that would produce only stars can
therefore be transformed to a MF with the correct shape and with BDs
and FFLOPs. The above estimate is very crude, and the pre-stellar
clump MF observed by Motte, Andr\'e \& N\'eri (1998) in $\rho$~Oph
which does not host an O~star is not a single power-law form but
already has the shape of the standard IMF, challenging the possibility
that photo-evaporation may be an important BD production channel.  The
result obtained here by simply setting the few parameters to
reasonable values, demonstrates above all else that photo-evaporation
of accretion envelopes may be an important process acting to shape the
IMF near and below the sub-stellar boundary within the immediate
vicinity of ionising stars.

Photo-evaporation would also affect forming binary systems by the
removal of binding mass from the proto-binary. The result (widened
binary or complete disruption) depends on the amount and rate of mass
loss (e.g. Hut \& Verhulst 1981), and is not considered further
here. It is only pointed out that this is a mechanism that would lead
to systematic changes in binary properties with decreasing primary
mass in the sense of producing wider and fewer binaries with
decreasing primary mass. The results of Close et al. (2003), Gizis et
al. (2003), Bouy et al. (2003) and Mart{\'{\i}}n et al. (2003) for
very-low-mass stars and for BDs in the Galactic field and the Pleiades
do not support this, however, because their separation distribution
appears to be very narrow and truncated near 20~AU, with a maximum
near 4~AU, unless most of the field BDs do not stem from ONC-type
clusters.

Photo-evaporation has an important kinematical implication that may be
used to test it. Only those proto-stars will loose substantial mass
from their accretion envelope that spend sufficient time near the
dominating O~star.  Since most young clusters have their O~stars
located near their centres, this may transform into a kinematical
bias: only those accreting proto-stars that have a low-enough velocity
dispersion will stay long enough near the cluster centre to have a
significant fraction of mass removed that would otherwise have ended
up in the star.

A population of accreting proto-stars that spend a time $\delta t$
confined within a radius $R_{\rm BD}$ has a velocity dispersion
$\sigma_{\rm BD} \approx R_{\rm BD} / \delta t $.  The bulk velocity
dispersion of the cluster which is assumed to be in virial equilibrium
before gas expulsion (Kroupa \& Boily 2003) is $\sigma_{\rm cl}^2
\approx G\, M_{\rm ecl+g} / 2\, R_{0.5}$, so that
\begin{equation}
{\sigma_{\rm BD}  \over \sigma_{\rm cl}} = {1\over \sqrt{\epsilon}}\,
                  {R_{\rm BD}\over \delta t}\,
                  \left({2\,R_{0.5} \over G\,M_{\rm ecl}}\right)^{1/2}.
\label{eq:relvd}
\end{equation}
The stellar mass of the cluster, $M_{\rm ecl}=M_{\rm ecl+g}/\epsilon$ for
a star-formation efficiency, $\epsilon$, of 33~per cent, can be
related simply to the mass of the most massive star, $m_{\rm u}$, by
assuming the IMF is a Salpeter power-law and insisting that there is
only one star with mass $m_{\rm u}$: $M_{\rm ecl} = \int_{0.1}^{m_{\rm
u}}\,m\,\xi(m)\,dm$, $\xi(m)=k\,m^{-2.3}$, $1=\int_{m_{\rm
u}}^{\infty}\,\xi(m)\,dm$. With a mass-luminosity relation for massive
stars of the form $L_{\rm u}/L_\odot = 1.2\,\left(m_{\rm
u}/M_\odot\right)^{3.8}$ (Cox 2000, p.~382) the following expression
results,
\begin{equation}
{M_{\rm ecl} \over M_\odot} = 
      8.24\, \left({L_{\rm u} \over L_\odot}\right)^{0.355} - 
      2.73\, \left({L_{\rm u} \over L_\odot}\right)^{0.263}.
\label{eq:Mcl}
\end{equation}

To illustrate the kind of kinematical effect one may expect it is
useful to take $R_{\rm BD}$ to be that radius within which all $m_{\rm
ps}=0.3\,M_\odot$ proto-stars are converted to $0.08\,M_\odot$
proto-stars within a time $\delta t = 0.1$~Myr. From eq.~\ref{eq:deltaM}
follows, with the condition $\delta m \le m_{\rm ps}$,
\begin{equation}
R = \left({\kappa \, L_{\rm u} \, \delta t \, R_{\rm ps}^3 \over
             4\,G\,m_{\rm ps}^2}\right)^{1/2} \, 
             \left( 2\,\left({\delta m \over m_{\rm ps}} \right) \,
	       - \left( {\delta m \over m_{\rm ps}} \right)^2 \right)^{-1/2}.
\label{eq:RBD}
\end{equation}
With $m_{\rm ps}=0.3\,M_\odot$ and $\delta m = 0.22\,M_\odot$ 
an estimate for $R_{\rm BD}=R$ results. 

\begin{figure}
\begin{center}
\rotatebox{0}{\resizebox{0.6 \textwidth}{!}
{\includegraphics{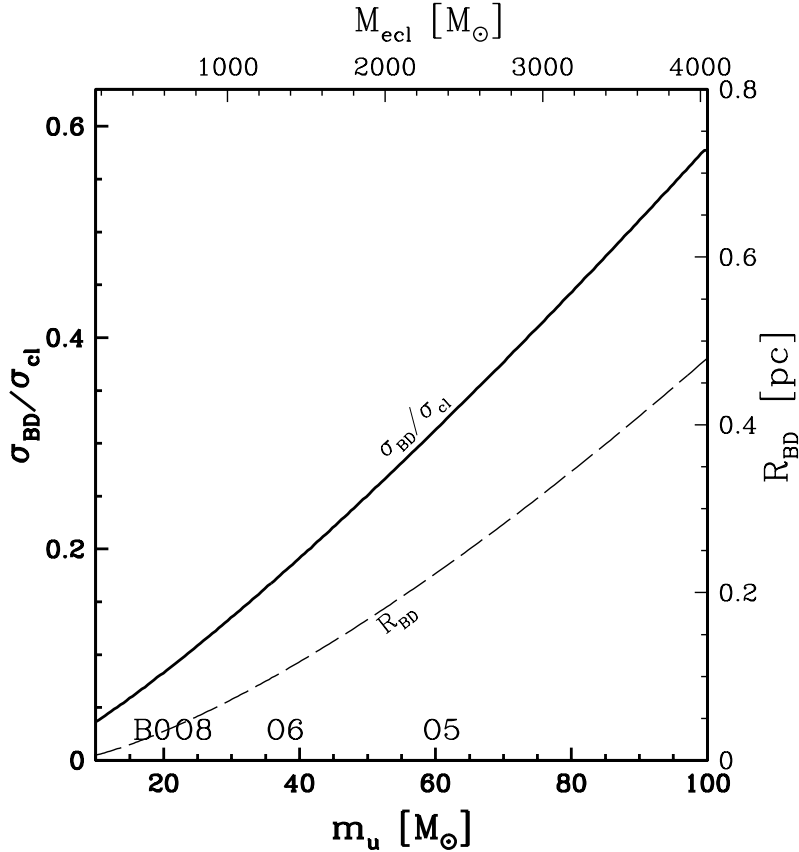}}}
\vskip -20mm
\caption
{The ratio of the velocity dispersion of the emerging BD and FFLOP
population to the stellar velocity dispersion in the cluster as a
function of the mass, $m_{\rm u}$, of the central ionising star (solid
line). The corresponding stellar mass of the cluster, $M_{\rm ecl}$,
is given on the top axis assuming the IMF is a Salpeter power-law. The
radius, $R_{\rm BD}$, within which a proto-star initially with a mass
of $0.3\,M_\odot$ is reduced to one with a mass of $0.08\,M_\odot$ is
shown as the dashed curve with corresponding axis on the right.  The
model assumes $\delta t=0.1$~Myr, $\kappa=0.1$, $R_{0.5}=0.4$~pc,
$m_{\rm ps}=0.3\,M_\odot$, $\delta m=0.22\,M_\odot$.  For example, a
central O5 star will transform all $<0.3\,M_\odot$ proto-stars to
sub-stellar objects if they remain within a radius of $R_{\rm
BD}\approx 0.2$~pc for 0.1~Myr.  This population of BDs and FFLOPs has
a velocity dispersion $\sigma_{\rm BD}\approx 0.3\,\sigma_{\rm cl}$.
The data plotted here are to be understood as illustrative of the type
of effect that we may expect (kinematically ``cold'' populations of
BDs centred about ``their'' massive stars in rich young clusters). }
\label{fig:vd}
\end{center}
\end{figure}

Fig.~\ref{fig:vd} illustrates the effect for a cluster with half-mass
radius $R_{0.5}=0.4$~pc and for different central photo-ionising
stars. The figure shows that the BD and FFLOP population should be
kinematically colder than the stars ($\sigma_{\rm BD} < \sigma_{\rm
cl}$), and that the effect may be more pronounced for less massive
clusters.  However, this only applies if most massive stars form near
the centre of their cluster. {  Should the massive stars form within
sub-clusters throughout the volume of the emerging cluster and then
sink to the centre through dynamical mass segregation then the
associated photo-evaporated BDs will mix with the cluster population
and acquire more or less virial velocities by the time the
sub-structure has been erased through the merging process.}

{  Given the numbers presented in Fig.~\ref{fig:vd} it is possible
to estimate which fraction of the stellar population could be
photo-evaporated BDs.  In the ONC that has one O6 star, $R_{\rm BD}
\approx 0.1$~pc, which is much smaller than the half-mass radius of
the pre-gas-expulsion cluster, $R_{0.5,{\rm ONC}}\approx 0.4$~pc, the
fraction of the stellar population that are photo-evaporated BDs is
${\cal R}_{\rm env}=0.007$, assuming the cluster had a Plummer density
profile (KAH) and if all M~stars within $R_{\rm BD}$ were to transform
to BDs (M~dwarfs contribute roughly 50~per cent to a stellar
population, Kroupa 2002).  This estimate of ${\cal R}_{\rm env}$ is
valid for the embedded cluster. Once the gas is removed and the
cluster expands the kinematically colder photo-evaporated BD
population will expand less than the kinematically hotter stellar plus
ejected-BD component, so that an observer will today find a larger
number of photo-evaporated BDs per star in the central region of the
ONC. An improved estimate of the contribution by photo-evaporated BDs
awaits $N-$body computations as energy equipartition between the BDs
and stars is likely to affect the results.}

{  The above estimate for $R_{\rm env}$ assumes that the ONC had
only one O~6 star ($\theta^1$~Ori~C). Photo-evaporation could have
been more effective should the initial cluster core have contained a
larger number of massive stars. This is possible because the presently
observed Trapezium configuration at the cluster centre is dynamically
highly unstable. Most of the Trapezium stars are higher-order multiple
systems (Preibisch et al 1999) that are interacting with each other on
a time-scale given by the dynamical time of the core of massive stars.
Thus, assuming the core had a dimension $R_{\rm
core}=0.05$~pc~$\approx 10^4$~AU and a mass $M_{\rm core}\approx
150\,M_\odot$ it should decay on a time-scale of approximately
$4\times 10^4$~yr (eq.~\ref{eq:tdyn}). Explosive gas expulsion
together with a low star formation efficiency in the core would also
expand it and allow massive stars to leave the central region (Vine \&
Bonnell 2003).  The very existence of the Trapezium at the present
time therefore poses an important challenge for understanding the
ONC. One possible solution is that the proto-ONC may have had a more
populated cluster core of which we today only see one snap-shot before
it decays further.  Also, the Trapezium may actually be rather young,
as evidenced by the short photo-evaporation time-scale of stellar
envelopes and the apparent youth of the dynamical configuration of the
ONC (Kroupa et al. 1999).}

{  The photo-evaporation model predicts the stellar IMF to vary with
conditions, since it implies that in dense and populous clusters some
or many of the M~dwarfs ought to be failed G~dwarfs.  Briceno et
al. (2002) and KBDM show that the stellar IMF in TA and the ONC are
very similar in the mass range $0.1-1\,M_\odot$, which ought not to be
the case if photo-evaporation of envelopes does play a major role.
Agreement of different IMFs in the stellar regime does not necessarily
preclude photo-evaporation being an important mechanism, because, in
essence (eq.~\ref{eq:deltaM}) $\delta m / m_{\rm ps,i} \propto m_{\rm
ps,i}^{-1}$ and because more massive proto-stars collapse to compact
morphologies more rapidly than less-massive proto-stars (this has not
been modelled here) so that the least-massive proto-stars are affected
significantly more than typical proto-stars. The emerging IMF will
therefore be most sensitive to the presence of O~stars in the
sub-stellar mass regime.}

{  For very massive clusters (stellar super-clusters), which may contain
many stars with masses reaching to $150\,M_\odot$, the radius for
producing photo-evaporated BDs from low-mass proto-stars may be
similar to the radius of the whole cluster, $R_{\rm BD}\approx R_{\rm
cl}$. {\it Globular clusters may thus have a significant population of
photo-evaporated BDs}.  The globular cluster stellar MFs appear to be
rather similar to the stellar MFs in young Galactic clusters (see
Kroupa 2002 for a review), although there is some tentative evidence
that young Galactic clusters may have an IMF decreasing more steeply
with increasing stellar mass (Kroupa 2001), thus allowing some
constraints on the degree with which photo-evaporation may affect the
stellar IMF in clusters that still contain or in the past contained
O~stars.}

\section{The collision model}
\label{sec:rem}

In dense clusters accreting proto-stars can pass each other on
hyperbolic orbits such that they loose a part of their accretion
envelope during the collision, the embryos being removed from their
gas reservoir (Price \& Podsiadlowski 1995).  {  The mechanism for
accretion-envelope removal envisioned here is different to the
embryo-ejection model because previously unbound proto-stars collide
with hyperbolic velocities whereby the envelopes interact
hydrodynamically thus dissipating the relative kinetic energy, while
the compact hydrostatic cores continue their fly-by largely
unhindered.}  The so produced BDs or FFLOPs remain trapped in the
cluster potential.  {\it The primary difference to the embryo-ejection
model is that the velocity dispersion of the BDs is the same as that
of the stars, while embryo ejection leads to a larger BD velocity
dispersion in the cluster}.  Any such destructive encounter will
disrupt the proto-binary within which the proto-star was accreting.
Thus, in the ONC some of the accreting BD companions may have been
lost from their parent systems and, having been deprived from gas to
accrete, they may have remained sub-stellar.

In TA stellar-dynamical encounters are rare (Kroupa \& Bouvier 2003),
and the accreting hydrostatic cores remain companions locked to the
stars.  They can continue to accrete for 1~Myr (or longer) since they
do not experience close encounters.  It would require a mass accretion
rate on the BD companion of only about $\dot{m} = 10^{-7}\,M_\odot$/yr
during the first Myr for it to become a star. This is much larger than
the currently estimated $\dot{m}$ for young BDs ($\approx 10^{-9}\,
M_\odot$/yr) but, like for stars, it is likely that the mass accretion
rate rapidly decreases with time. It is typically
$10^{-8}\,M_\odot$/yr for a 2 Myr T~Tauri star, but is believed to be
as high as $10^{-6}\,M_\odot$/yr during the embedded proto-stellar
phase which lasts several $10^5$~yr.  Initial BD companions to stars
in TA could therefore become (low-mass) stars in about 1~Myr (fig.~1
in Reipurth \& Clarke 2001).  This could account both for the
deficiency of isolated BDs in TA compared to the ONC and for the
period distribution of stellar binaries in TA, since most/all
potential BD companions would have evolved into stellar companions.
This is probably the physical reason why the star-like BD flavour had
to be excluded in \S~\ref{sec:intro} as contributing a significant
fraction of the overall BD population.

Assuming the same IMF in TA and the ONC which is truncated at the
hydrogen burning mass limit (as in \S~\ref{sec:phot}) we want to
investigate if the collision process may be relevant for the observed
different number of BDs per star (and FFLOPs) in the ONC and TA.

The role of collision-induced accretion-envelope removal has been
studied in detail by Price \& Podsiadlowski (1995) for a range of
environments. The general result of their work is that the stellar
(and sub-stellar) IMF should vary in dependence of the richness of the
cluster. Their models can be applied to the case of the ONC to obtain
an indication of the type of effect that might occur. We note however,
that a number of parameters that describe the model IMF obtained in
this way are rather uncertain, such as the star-formation history, the
shape of the proto-stellar envelopes and mass-accretion rates.

The proto-stellar collision rate (per cluster member) is given by
their eq.~2.1,
\begin{equation}
\beta_{\rm pp}= {3\,R_{\rm ps}^2\, \sigma_{\rm cl} \over 4\,R_{\rm cl}^3},
\label{eq:bpp}
\end{equation}
where $R_{\rm cl}$ is the characteristic cluster radius. For the
proto-ONC $R_{\rm cl}\approx0.5$~pc and $\sigma_{\rm cl}\approx
5$~pc/Myr (KAH) so that $\beta_{\rm pp}\approx R_{\rm ps}[{\rm pc}]^2\
\times 3\times10^{-5}$~yr$^{-1}$ per member. For $R_{\rm ps}\le100$~AU
$=10^{-3.31}$~pc, $\beta_{\rm pp}\le10^{-11}$~yr$^{-1}$ per
member. Such a small $\beta_{\rm pp}$ implies a Salpeter IMF down to
about $0.01\,M_\odot$ (fig.~3b in Price \& Podsiadlowski 1995), which
is not observed. Also, over 0.1~Myr there are $\le 10^{-6}$ collisions
per member with impact parameters $\le 200$~AU. With $10^4$ ONC
members (KAH) we obtain $\le 10^{-2}$ possible BDs in the ONC.  This
is much smaller than the few hundred BDs found so far, indicating that
collisional processes are completely negligible in the ONC, as already
stressed in Kroupa (2002).

{  Scally \& Clarke (2001) perform $N-$body computations of an
ONC-type cluster to estimate the number of encounters that could have
truncated circum-stellar disks. Their model cluster consists of
4000~stars with a half-mass radius of 1~pc, and they find that by
2.89~Myr only about 4~per cent of the stars suffered encounters with
peri-astra less than 100~AU. The encounter rate is approximately
$\rho\,\pi\,R_{\rm ps}^2\,\sigma_{\rm cl}$, where $\rho$ is the
cluster number density.  Scaling this to our ONC model, the 4~per cent
thus needs to be multiplied by $(\rho/\rho_{\rm SC}) \times (R_{\rm
ps}/100\,{\rm AU})^2 \times (\sigma_{\rm cl}/\sigma_{\rm cl,SC})$,
where the values with sub-script SC refer to the Scally \& Clarke ONC
model. We have $\rho/\rho_{\rm SC} = (10000/0.5^3) / 4000 = 20$,
$\sigma_{\rm cl}/\sigma_{\rm cl,SC} = 7 / 4.3 = 1.6$. The ONC model
used here thus implies that about 1~per cent of all stars will have
experienced encounters with peri-astra less than 10~AU over 2.89~Myr,
or 0.04~per cent over a time span of 0.1~Myr, which is the life-time
of a proto-star before it evolves to a more compact star plus low-mass
disk morphology. For an ONC membership of $10^4$ this amounts to
4~possible collisional BDs confirming that collisional processes will
have been negligible in the ONC.}

Overall therefore, {\it the evidence points strongly against
collisional effects being an important channel for producing BDs}.

\section{A combined model}
\label{sec:comb}

{  In \S~\ref{sec:ej} a successful model was developed that accounts for
the observed number of BDs per star in the TA and the ONC by assuming
that BDs come in only one flavour, namely the embryo-ejected type.
However, \S~\ref{sec:phot} demonstrated that photo-evaporated BDs may
also constitute an existing BD flavour in clusters containing O~stars,
while star-like (\S~\ref{sec:intro}) and collisional
(\S~\ref{sec:rem}) BDs appear to be very rare.}

{  The stars in the ONC are faster rotators than in TA (Clarke \&
Bouvier 2000), and this may be another indicator that some stars may
have lost their accretion envelopes earlier in the ONC thus reducing
disk breaking.  The prediction of the photo-evaporation (and
collision) scenario is that {\it the binary proportion should decrease
with the rotational velocity of the stars}. That is, if the ONC
late-type stars are split into fast and slow rotators, then the fast
rotators should have a smaller binary fraction than the slow rotators
because they may have suffered mass-loss through photo-evaporation or
an encounter.}

{  In this section we consider all four BD flavours to investigate
if a consistent description of the observations can be attained (the
star-like and collisional BDs are kept only formally).  Here the
ansatz is to again assume that the ejected embryos have the same
distribution function of ejection velocities in TA and the ONC
(eqs~\ref{eq:schw} and~\ref{eq:sigma}): that in TA and the ONC the
ejections occur from dynamically unstable small$-N$ systems that have
the same characteristics in both star-forming regions. As in
\S~\ref{sec:ej} ejected BDs can be lost from the cluster (ONC) or
stellar group (TA), while photo-evaporated (and star-like and
collisional) embryos are retained in the cluster or group.}

For a BD and FFLOP population that consists of ejected,
photo-evaporated and collisional embryos an observer will detect
(cf. eq,~\ref{eq:R3})
\begin{equation}
{\cal R}_{\rm obs} = (1+f)\,\left[{\cal R}_{\rm ej} \,  
               \left({\cal B} + {\cal U}\right) + 
               {\cal R}_{\rm o}\right]
\label{eq:comb}
\end{equation}
BDs per stellar system, where ${\cal R}_{\rm o}={\cal R}_{\rm env} +
{\cal R}_{\rm st}$ and ${\cal R}_{\rm env} = {\cal R}_{\rm phot}+{\cal
R}_{\rm coll}$.  Here ${\cal R}_{\rm st}\equiv N_{\rm BD,st}/N_{\rm
st,tot}$ is the number of star-like BDs per star, ${\cal R}_{\rm
phot}\equiv N_{\rm BD,phot}/N_{\rm st,tot}$ is the number of
photo-evaporated BDs per star, ${\cal R}_{\rm coll}\equiv N_{\rm
BD,coll}/N_{\rm st,tot}$ is the number of collisional BDs per star,
and ${\cal R}_{\rm ej}\equiv N_{\rm BD,ej}/N_{\rm st,tot}$ is the
number of BDs per star produced by embryo ejection, while ${\cal
B}(\sigma_{\rm ej,1D},M_{\rm ecl+g})$ is the fraction of ejected BDs
retained in the cluster or group potential, and ${\cal U}(\sigma_{\rm
ej,1D},M_{\rm ecl+g})$ is the fraction of ejected and unbound BDs
which are still detected within the observational survey region
because they haven't drifted far enough by the time of observation.
Finally, the evidence points to ${\cal R}_{\rm st}\approx 0$
(\S~\ref{sec:intro}) and ${\cal R}_{\rm coll}\approx0$
(\S~\ref{sec:rem}).

\begin{figure}
\begin{center}
\rotatebox{0}{\resizebox{0.6 \textwidth}{!}
{\includegraphics{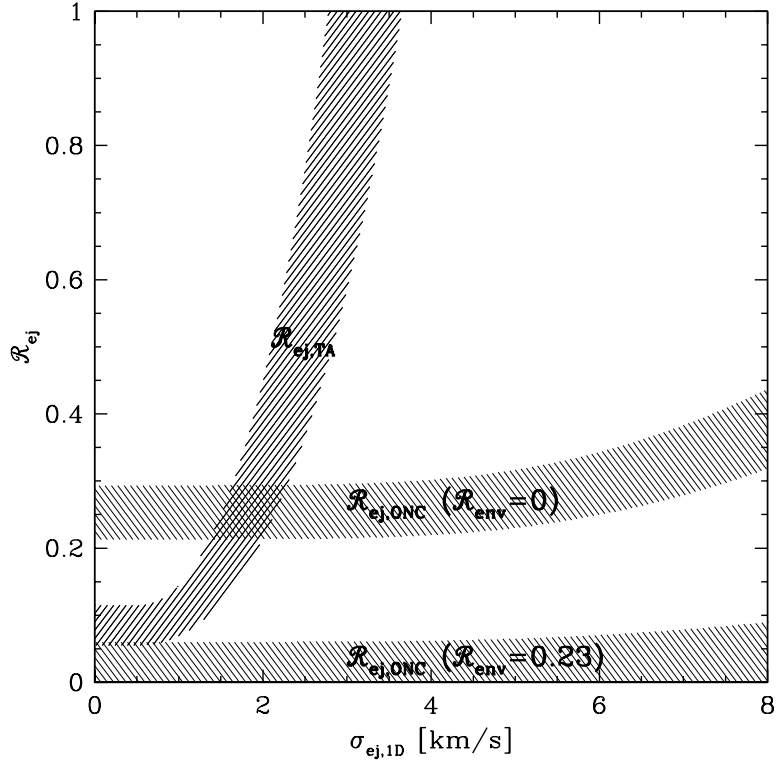}}}
\vskip -25mm
\caption
{The number of ejected BDs per star (${\cal R}_{\rm ej}$) in TA and
the ONC as a function of the velocity dispersion of the ejected
population.  If accretion envelope removal through photo-evaporation
or stellar-dynamical collisions plays no role in creating BDs and
FFLOPs from embryos that would otherwise become stars (${\cal R}_{\rm
env}=0$) then about one BD per four stars is produced through ejection
of unfinished embryos with $\sigma_{\rm ej,1D}\approx 2$~km/s in TA
and the ONC (recovering the result from \S~\ref{sec:ej}). If, on the
other hand, ${\cal R}_{\rm env}=0.23$ photo-evaporated BDs per star in
the ONC, then the observational data require the rate of embryo
ejection to be reduced to~6 BDs per 100 stars (${\cal R}_{\rm
ej}\approx 0.06$), and $\sigma_{\rm ej,1D}< 0.8$~km/s in TA and the
ONC. Note that in all cases $\delta=0$ (eq.~\ref{eq:delta}).  }
\label{fig:comb}
\end{center}
\end{figure}

To assess what minimum $R_{\rm ej}$ can give rise to the observed
values in TA (${\cal R}_{\rm obs,TA}=0.17\pm0.06$) and the ONC (${\cal
R}_{\rm obs,ONC}=0.38\pm0.06$) for the same distribution of ejection
velocities, eq.~\ref{eq:comb} is solved for ${\cal R}_{\rm ej}$ to
obtain (${\cal R}_{\rm env,TA}=0$)
\begin{eqnarray}
{\cal R}_{\rm ej,TA} &=& {{\cal R}_{\rm obs,TA} \over 
             1+f_{\rm TA}} \, 
             \left({\cal B}_{\rm TA}+{\cal U}_{\rm TA}\right)^{-1}, 
             \quad\quad{\rm and} \nonumber\\
{\cal R}_{\rm ej,ONC} &=& 
         \left({{\cal R}_{\rm obs,ONC} \over 1+f_{\rm ONC}}
         -{\cal R}_{\rm env}\right) 
             \, \left({\cal B}_{\rm ONC}+{\cal U}_{\rm ONC}\right)^{-1}.
\label{eq:rej}
\end{eqnarray}
Note that ${\cal U}_{\rm ONC}=0$ for $\sigma_{\rm ej,1D}<8$~km/s
(Fig.~\ref{fig:fr}).  The result is displayed in Fig.~\ref{fig:comb}
for ${\cal R}_{\rm env}=0$ and ${\cal R}_{\rm env}=0.23$.  The result
from \S~\ref{sec:ej} is arrived at again: ${\cal R}_{\rm TA}={\cal
R}_{\rm ONC}=0.21-0.29$ for $\sigma_{\rm ej,1D}\approx2$~km/s.
However, if about 0.23~BDs per star are produced through
photo-evaporative envelope loss in the ONC, then ${\cal R}_{\rm
TA}={\cal R}_{\rm ONC} \approx 0.06$ for $\sigma_{\rm ej,1D}\simless
0.8$~km/s. In this case a large fraction of ejected BDs are retained
in both, the TA groups and in the ONC (${\cal B}_{\rm TA} \simgreat
0.5; {\cal B}_{\rm ONC} \approx 1$, Fig.~\ref{fig:fr}).

{  It therefore seems possible that BDs come mostly in two flavours,
namely as ejected embryos from all star forming regions and as
photo-evaporated embryos.  Present data for TA and the ONC allow
limits to be placed on the production rate (per star) of each flavour,
ranging from~0.21--0.29 ejected BDs per star if the other three
production channels can be neglected altogether, down to~0.06 ejected
BDs per star if photo-evaporated BDs contribute 0.23~BDs per star in
clusters similar to the ONC.}

\section{Does brown-dwarf birth depend on environmental conditions?}
\label{sec:fin}

As pointed out in \S~\ref{sec:ej} there appears to exist a
disconcerting discrepancy between the observed number of BDs per star
in TA and the ONC and the observed number of BDs in the Galactic
field. The analysis performed above suggests that {\it there can be at
most ${\cal R}_{\rm ej}\approx 0.25\pm0.04$ BDs per star produced as
ejected embryos with a velocity dispersion of about 2~km/s if
photo-evaporated BDs can be neglected}.  If, on the other hand, ${\cal
R}_{\rm env}=0.23$ BDs per star are embryos that lost their accretion
envelope due to photo-evaporation in ONC-type clusters, then only
${\cal R}_{\rm ej}\approx0.06$ BDs per star are created as ejected
embryos in all environments with a small velocity dispersion of only
about 0.8~km/ or less.

In young clusters similar to the ONC the maximum number of BDs per
star is thus~${\cal R}=0.25+0.12=0.37$ (three-sigma limit).  This is
significantly smaller than the number of BDs per star expected from
the Galactic-field MF (eq.~\ref{eq:Rimf}, ${\cal R}=0.81$ for
$\alpha_0=+0.3$, Kroupa 2002; ${\cal R}\approx 1$, Chabrier 2002),
unless the surveys lead to an overestimate of the number of BDs per
star in the Galactic field. 

This problem may be remedied by dropping the ansatz made in the above
analysis that the ejected-embryo population have the same distribution
of ejection velocities, independent of the star-formation conditions,
and that the number of ejected BDs per star produced be the same. { 
That is, we now allow the physical properties of the fragmenting cloud
kernels to vary with the conditions in their molecular cloud.}

Thus, from Fig.~\ref{fig:comb}, ${\cal R}_{\rm ej,TA}=1$ is consistent
with the observational datum for TA if $\sigma_{\rm
ej,1D}\approx3$~km/s.  The weak potentials of the TA groups cannot
hold most of the ejected BDs, and {\it the prediction would be that a
large fraction of the ejected BDs are distributed throughout the
star-forming area but outside the surveyed groups}, ${\cal B}_{\rm TA}
\approx 0.05$ in Fig.~\ref{fig:fr}, and a BD with a velocity of 3~km/s
leaves the Briceno et al. (2002) survey area, 2.3~pc x 2.3~pc, within
less than 1~Myr. On the other hand, the ONC has never produced the
number of BDs per star evident in the Galactic field, unless the
dispersion of velocities was unreasonably high (Fig.~\ref{fig:comb}:
$\sigma_{\rm ej,1D} > 10$~km/s for ${\cal R}_{\rm ej,ONC}\approx1$).
The potential well of the ONC is so deep that virtually all BDs are
kept for reasonable ejection velocity-dispersions, and the
observational datum, ${\cal R}_{\rm obs,ONC}$, forces the number of
ejected BDs per star to lie near~$0.25\pm0.04$ for $\sigma_{\rm
ej,1D}\simless 5$~km/s (${\cal B}_{\rm ONC}>0.9$,
Fig.~\ref{fig:fr}). {\it Adding other BDs flavours reduces the
possible contribution by ejected BDs even further, and it appears thus
that most Galactic-field BDs may not have been born in ONC-type
clusters}, unless the number of Galactic-field BDs is overestimated.

{  This interpretation of the data would imply that cloud kernels in
TA can fragment more vigorously thereby ejecting many more unfinished
embryos than in the early ONC.} However, this would appear to
contradict the Jeans-mass argument (\S~\ref{sec:intro}).  An unbiased
determination of ${\cal R}$ in the solar-neighbourhood is not a
trivial matter given the various uncertainties (unknown ages, masses
and distances) that still affect the census of field BDs so that the
solutions presented in \S~\ref{sec:ej} and \S~\ref{sec:comb} that rely
on a universal ejection process are probably to be favoured.

\section{Concluding remarks}
\label{sec:concs}

Recent findings on BD properties are now allowing interesting insights
into the nature and origin of BDs. The observational surveys by Close
et al. (2003), Gizis et al. (2003), Bouy et al. (2003) and
Mart{\'{\i}}n et al. (2003), and the study by KBDM suggest that BDs
probably did not form with the same properties as stars (i.e. as
hydrostatic cores that accrete most of their available envelope mass),
because the predicted properties of star--BD and BD--BD binaries are
not consistent with available constraints. The star--BD and BD--BD
binaries should have properties that are a natural extension of the
trends seen among stars with decreasing primary mass, which does not
appear to be the case as emphasised by Close et al. Most BDs therefore
do not appear to have a star-like accretion history. Instead, the
majority of BDs probably start-off like stars but their accretion is
truncated through an external agent. The aim of this study is to
investigate which implications the three possible truncation
mechanisms (embryo ejection, photo-evaporation and hyperbolic
collisions) may have on the observable properties and distribution of
BDs.

The study performed here shows that the embryo-ejection hypothesis can
lead to a consistent description of the number of BDs per star seen by
Briceno et al. (2002) in the TA groups and in the central region of
the ONC.  Owing to the weaker gravitational field, TA groups retain a
small fraction of their BDs, while the ONC captures most of them.
{\it In both environments about one BD is ejected from a multiple
system containing on average four stars (${\cal R}_{\rm ej}\approx
0.21-0.29$)}. According to this picture the fragmenting cloud kernels
would have, statistically, the same physical properties in TA and the
ONC on scales less than a few hundred~AU.  The different stellar
binary fraction in TA and in the ONC is taken into account explicitly
in the calculation of $R_{\rm ej}$.  The dispersion of ejection
velocities comes out to be about~2~km/s, and this poses the presently
simplest and therefore favoured interpretation of the data. The
rejection of the embryo-ejection hypothesis by other workers
(\S~\ref{sec:intro}) may not be valid because ejected BDs can retain
disks (\S~\ref{sec:ej}) and because the detected BDs have $v<v_{\rm
in}$, as otherwise they would not appear in the observed volumes,
implying similar kinematics to the stars. Larger survey areas are
needed to better address this latter point. An interesting possibility
related to the embryo-ejection hypothesis is that the sharp rise of
the MF observed by Muench et al. (2002, 2003) in the ONC and the
IC~348 below $0.03\,M_\odot$ may be due to the rapid dynamical decay
of young many-planet systems.

In rich clusters that contain O~stars photo-evaporation of the
accretion envelope of low-mass hydrostatic cores, that would otherwise
become M~dwarfs, can instead produce a BD or a FFLOP. With this
mechanism, a featureless Salpeter power-law proto-stellar MF, that
would produce only stars in the absence of O~stars, can transform to
the observed Galactic-field IMF which has a flattening near
$0.5\,M_\odot$ and which contains BDs and FFLOPs. {\it A population of
photo-evaporated BDs should have a smaller velocity dispersion than
the stars, with a trend to smaller relative velocity dispersion for
less massive clusters}. Photo-evaporated BDs should be confined to a
small vicinity about the ionising star.  Both of these observable
diagnostics apply unless massive stars form with their own
sub-clusters throughout the volume encompassing the emerging cluster,
because the BD sub-populations will mix and virialise within the
emerging cluster.  In very massive stellar super clusters that contain
many stars as massive as $150\,M_\odot$ photo-evaporated BDs probably
contribute significantly to the BD population.  That the removal of
accretion envelopes may have been occurring in the ONC is supported
tentatively by the late-type stars in the ONC being faster rotators
than in TA.

BDs and FFLOPs can also be produced through removal of accretion
envelopes due to hyperbolic collisions of $\simless 0.1$~Myr old
proto-stars. The collision scenario predicts the binary fraction to
decrease with increasing rotational velocity
(\S~\ref{sec:comb}). However, the expected rate of collisions is
negligible in the ONC.

If the ansatz is retained that the physical properties of the
fragmenting cloud kernels and thus the number of ejected BDs per star
be the same independent of environment, then adding photo-evaporated
BDs into the ONC reduces the allowed number of BDs that could have
been produced in the TA groups and in the ONC as ejected embryos, from
about one BD per four stars (in the absence of the other BD flavours)
down to about six ejected BDs per 100~stars (${\cal R}_{\rm ej}\approx
0.06$) with a small dispersion of ejection velocities of $\sigma_{\rm
ej,1D}\simless 0.8$~km/s if there are 0.23~photo-evaporated BDs per
star in the central region of the ONC.

The ansatz that $R_{\rm ej}$ and $\sigma_{\rm ej,1D}$ be the same in
TA and in the ONC leads to the uncomfortable situation that the number
of BDs per star (maximally ${\cal R} = 0.37$, three-sigma value) is
inconsistent with independent measurements of the BD MF in the
Galactic field (${\cal R}\approx 0.9$). Discarding this ansatz, it is
found that only in TA can star formation have produced about one BD
per star (${\cal R}_{\rm ej,TA}\approx 1$) if $\sigma_{\rm
ej,1D}\approx 3$~km/s. The observational data from the ONC do not
allow one~BD to have been produced per star in the proto-ONC. Instead,
star formation in the ONC could only have produced at most 0.37~BDs
per star either wholly as ejected embryos, or together with
photo-evaporated BDs.  {\it If the inferred number of BDs per star in
the Galactic-field is not an overestimate, then this indicates that
the number of BDs per star may depend sensitively on the star-forming
conditions in the sense that low-mass tranquil star-formation may be
producing most of the known BDs, while ONC-type clusters may be
relatively inefficient in producing BDs.}  Note that this conclusion
would be opposite to an unreflected interpretation of the TA and ONC
data (first sentence in the abstract) and would also be inconsistent
with the Jeans-mass argument (\S~\ref{sec:intro}).

The ONC datum used in the analysis presented here is based on the
number of BDs per star in the central cluster region. Preibisch et
al. (2003) find that this datum is consistent with the independent
surveys by Muench et al. (2002) for the central region, and for the
whole-ONC survey by Hillenbrand \& Carpenter (2000). Preibisch et
al. (2003) and Luhman et al. (2003) measure the BD content of IC~348
and find this cluster, which does not contain O~stars and is
intermediate in density between the TA groups and the ONC, to be
similarly deficient in BDs (per star) as the TA groups.  According to
Preibisch et al. this may be due to most of the BDs in the ONC being
of the photo-evaporated flavour, but they admit that this may be
problematical since in the ONC significant photo-evaporation is
limited to a relatively small radius around $\theta^1$~Ori~C. As
suggested here, the ONC may have contained a larger number of massive
stars that were expelled from a dynamically unstable cluster core, so
that exclusion of photo-evaporation as the mechanism of providing a
substantial number of BDs in the ONC may be premature. Preibisch et
al.  also point out that the embryo-ejection hypothesis may well be
the main origin of most BDs, the ONC being much more efficient in
capturing the BDs than the much less massive IC~348 cluster and the TA
groups.

This work has found that embryo ejection is probably the main channel
for producing BDs, but the photo-evaporated flavour probably also
exists in rich clusters.  The BD situation remains exciting and
partially controversial; additional observational data are badly
needed to help refine our understanding of the origin and nature of
BDs and FFLOPs. Such data would be a census of BDs outside the stellar
groups in TA, improved constraints on the binary properties of BD
primaries, verification of the reported dramatic rise of the BD (or
FFLOP) MF below $0.03\,M_\odot$, as well as measurements of the
velocity dispersion and rotational velocities of BDs and stars in
clusters.

\section*{acknowledgements} 
PK thanks the staff of the Observatoire de Grenoble for their very
kind hospitality and the Universit\'e Joseph Fourier for supporting a
very enjoyable and productive stay during the summer of 2002, and
acknowledges partial support through DFG grant KR1635/4.


\vfill 

\end{document}